\documentclass[%
 reprint,
    superscriptaddress,
    amsmath,amssymb,
    aps,
]{revtex4-2}

\usepackage{graphicx}
\usepackage{dcolumn}
\usepackage{bm}
\usepackage{physics}

\begin{document}

\title{Two-center Interference in the Photoionization Delays of Kr$_2$}%

\author{Saijoscha Heck}
\author{Meng Han} 
\email{menhan@ethz.ch}
\author{Denis Jelovina}%
\author{Jia-Bao Ji}
\author{Conaill Perry}
\affiliation{ Laboratorium f\"{u}r Physikalische Chemie, ETH Z\"{u}rich, 8093 Z\"{u}rich, Switzerland.}
\author{Xiaochun Gong}
\affiliation{State Key Laboratory of Precision Spectroscopy, East China Normal University, Shanghai, China}
\author{Robert Lucchese}
\affiliation{Lawrence Berkeley National Laboratory, Berkeley, California 94720, USA}
\author{Kiyoshi Ueda}
\affiliation{ Laboratorium f\"{u}r Physikalische Chemie, ETH Z\"{u}rich, 8093 Z\"{u}rich, Switzerland.}
\affiliation{Department of Chemistry, Tohoku University, Sendai, 980-8578, Japan
}
\author{Hans Jakob Wörner}
\email{hwoerner@ethz.ch}
\affiliation{ Laboratorium f\"{u}r Physikalische Chemie, ETH Z\"{u}rich, 8093 Z\"{u}rich, Switzerland.}

\date{\today}

\begin{abstract}
We present the experimental observation of two-center interference in the ionization time delays of Kr$_2$. Using attosecond electron-ion-coincidence spectroscopy, we simultaneously measure the photoionization delays of krypton monomer and dimer. The relative time delay is found to oscillate as a function of the electron kinetic energy, an effect that is traced back to constructive and destructive interference of the photoelectron wave packets that are emitted or scattered from the two atomic centers. Our interpretation of the experimental results is supported by solving the time-independent Schrödinger equation of a 1D double-well potential, as well as coupled-channel multiconfigurational quantum-scattering calculations of Kr$_2$. This work opens the door to the study of a broad class of quantum-interference effects in photoionization delays and demonstrates the potential of attosecond coincidence spectroscopy for studying weakly bound systems.
\end{abstract}

\maketitle

Two-center interference is one of the most prominent manifestations of the wave character of matter. The simplest demonstration consists of a double slit as it was first done in 1801 by Thomas Young with light waves \cite{Young1803} and in 1961 by Claus Jönsson with electrons \cite{Jonsson1961}. Soon after that it was noted by Cohen and Fano \cite{Cohen1966} that the electron wave from photoionization of diatomic molecules resembles the one behind the double slit. Since then, there have been numerous investigations of the molecular double slit in diatomic molecules \cite{Liu2006,Le2006,Akoury2007,Schoffler2008,Baker2008,Kreidi2008,Chen2008,Hargreaves2009,Chaluvadi2015,Li2018}. The interference can be simply described with the superposition of two spherical waves departing from each atom of a diatomic molecule:
\begin{equation}
\Psi_{1,2}=\frac{1}{|\bm{r}|}\cdot e^{i(\bm{k}(\bm{r}\pm \bm{R}/2)+\Phi)},
\end{equation}
with an internuclear distance $\bm{R}$, momentum $\bm{k}$ and initial phase shift $\Phi$ \cite{Kunitski2019}. So far, most of the experiments have studied the photoionization cross section of aligned \cite{Liu2006,Akoury2007,Kreidi2008,Schoffler2008,Hargreaves2009,Li2018,Nagy2004} and unaligned \cite{Cohen1966} diatomic molecules. More recently the influence of two-center interference on high-harmonic generation (HHG) was investigated in CO$_2$ \cite{Kanai2005,Vozzi2005,Le2006,Rupenyan2013b,Rupenyan2013a} and H$_2$ \cite{VanDerZwan2010,Baker2008}. 

Owing to the fact that photoionization delays are indeed closely linked with the variation in the cross section \cite{Dahlstrom2013}, it is expected that two-center interference also has a significant impact on the ionization dynamics in the time domain. Vladislav Serov and other theoretical physicists made several pioneering predictions of such effects \cite{Serov2012Single,Serov2012Attosecond,Serov2013Interpretation,serov2014p,Liao2021,Ning2014a,pazourek15a} on H$_2$ and H$_2^+$ molecules. However, until now there has been no experimental observation of the influence of the two-center interference on the photoionization delays. Here, we report the photoionization delay of the krypton dimer relative to its monomer and observe oscillations in the delay that can be traced back to the interference of the electron wave packets that are emitted or scattered from the two weakly bound atoms in Kr$_2$.

The experiment was performed by combining an XUV attosecond pulse train (APT) generated via high-harmonic generation (HHG) in a 3~mm long gas cell filled with 20~mbar of xenon, covering the odd-order harmonics from H9 to H21, with an electron-ion coincidence spectrometer. The APT is focused into a cold krypton gas beam which is produced via supersonic expansion, where it is spatially and temporally overlapped with a near-infrared (NIR) pulse of co-linear polarization. The APT and NIR pulses are phase locked in an actively-stabilized Mach-Zehnder interferometer and their delay is controlled with a piezo-electric translation stage. Upon photoionization, the electrons and ions are detected in coincidence using COLd Target Recoil Ion Momentum Spectroscopy (COLTRIMS) \cite{Ullrich2003a,Dorner2000a}, which measures the three-dimensional momentum vectors of electrons and ions. A more detailed account of the experimental apparatus can be found in \cite{Heck2021}. The photoelectron spectra of Kr and Kr$_2$ are measured simultaneously for XUV-NIR delays between 0 to 7~fs, using the Reconstruction of Attosecond Beating By two-photon Transitions (RABBIT) technique \cite{Paul2001a,kluender11a,huppert16a,gong21a}. In RABBIT the intensity of the sidebands, which are the photoelectron bands generated by the additional absorption or emission of a single NIR photon by a photoelectron, oscillates as a function of the XUV-NIR delay $\tau$ as
\begin{equation} \label{RABBIT}
	I_\text{SB}=A+B*\cos(2\omega_{\text{NIR}}\tau-\Phi_{\text{XUV}}-\Phi_{\text{sys}}),
\end{equation}
where A and B are constants, $\omega_\text{NIR}$ is the center frequency of NIR. $\Phi_{\text{XUV}}$ is the spectral phase difference between the two adjacent harmonic orders (which characterizes the attochirp) and $\Phi_{\text{sys}}$ is the system-specific phase term. The latter is what we are interested in.

\begin{figure}
\begin{center}
\includegraphics[width=8.6cm]{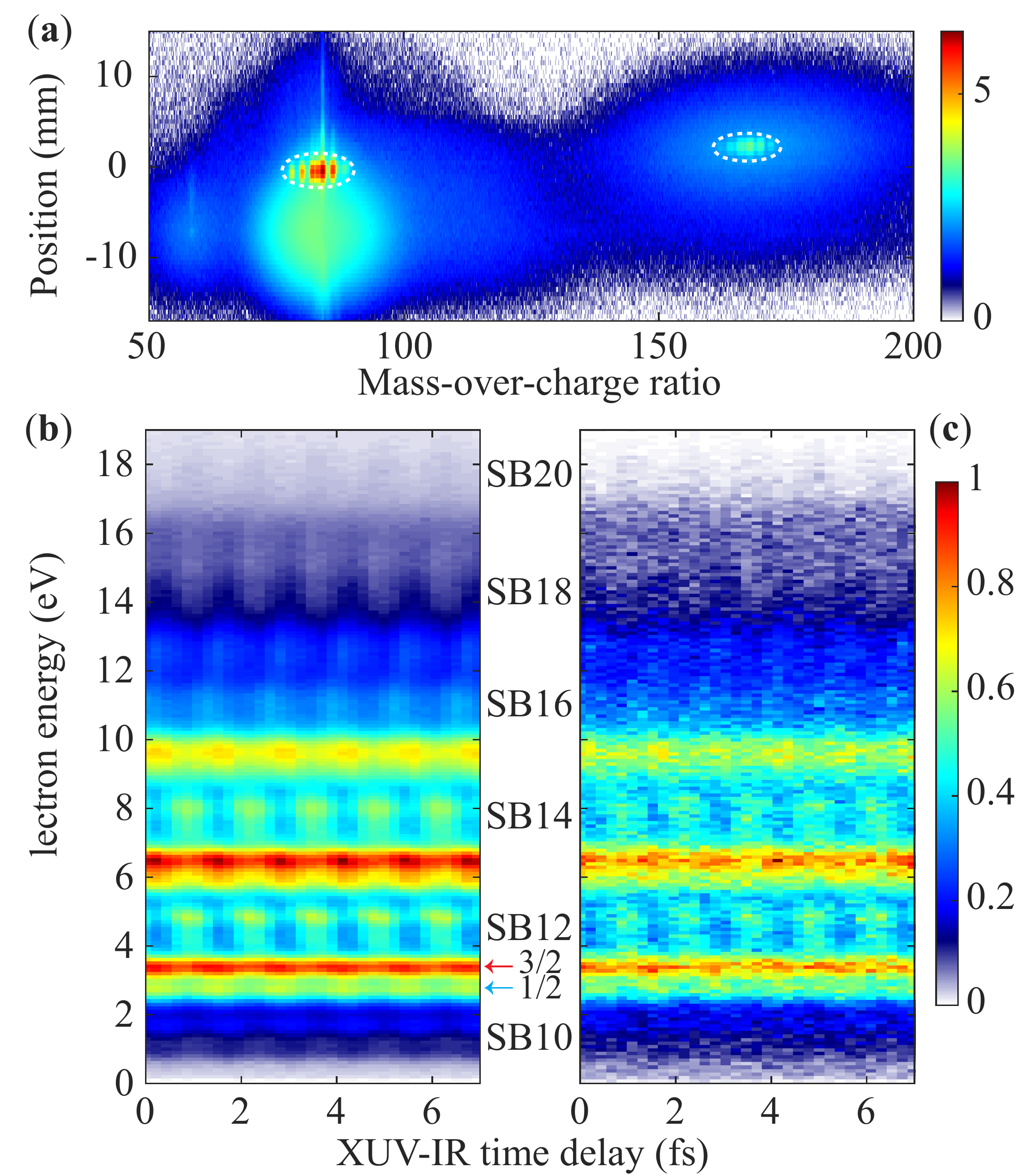}
\end{center}
{\textbf{Fig. 1.} (\textbf{a}) Measured ionic distribution as a function of the mass-over-charge ratio and the hit position on the detector along the direction of the supersonic molecular beam. The counts are shown on a logarithmic scale and displayed in false color. (\textbf{b}, \textbf{c}) RABBIT spectrograms for electrons detected in coincidence with undissociated Kr$^+$ and Kr$_2^+$, corresponding to the sharp distributions labeled with dashed ellipses in (a), respectively. Counts are normalized and shown in false color. In (b), the red and green arrows indicate the energy positions ionized by harmonic 11 for the two spin-orbit-coupling split states $^2$P$_{3/2}$ and $^2$P$_{1/2}$ of Kr$_2^+$, respectively.}
\label{figure1}
\end{figure}

In Figure 1a, we illustrate the measured ionic distribution as a function of the mass-over-charge ratio and the hit position on the detector. The sharp distributions of Kr$^+$ and Kr$_2^+$ (see dashed ellipses in Fig. 1a) indicate that these ions are from the undissociated channel, and their surrounding diffuse distribution of ions originates from the dissociative ionization channels of the larger clusters due to the kinetic-energy release in fragmentation. Figures 1b and 1c show the RABBIT spectrograms for photoelectrons measured in coincidence with the undissociated Kr$^+$ and Kr$_2^+$, respectively. The photoelectrons were detected for an emission cone angle of $\theta_{\rm{Lab}}=0-25^{\circ}$ between the electron momentum vector and the XUV polarization, where the molecular axis with respect to the XUV polarization is randomly oriented. Six sidebands ranging from SB10 to SB20 can clearly be seen in both spectra, as labeled. There is also the signature of spin-orbit splitting, which can be observed in Kr \cite{jordan17a}, as well as in Kr$_2$ (see the arrows in Fig. 1b). In the analysis of the sideband oscillations the energy range of each sideband was chosen to include both spin-orbit states. Due to the direct comparison of the same sidebands of Kr and Kr$_2$, the XUV spectral phase $\Phi_{\text{XUV}}$ cancels out and the relative photoionization delays are determined by
\begin{equation}
    \Delta \tau_{\rm Kr_2-Kr}=\hbar \frac{\partial \Phi_{\text{sys}}^{\rm Kr_2}}{\partial E} -\hbar \frac{\partial \Phi_{\text{sys}}^{\rm Kr}}{\partial E} \simeq \hbar \frac{\Delta \Phi_{\text{sys}}^{\rm Kr_2} - \Delta \Phi_{\text{sys}}^{\rm Kr}}{\Delta E},
\end{equation}
where $\Delta E =  2 \hbar \omega_{\text{NIR}}$ is the oscillation frequency of the sidebands. The most fundamental difference between monomer and dimer is that the two-center potential of the dimer will cause additional effects on its photoionization delay. 

\begin{figure}
\begin{center}
\includegraphics[width=8.6cm]{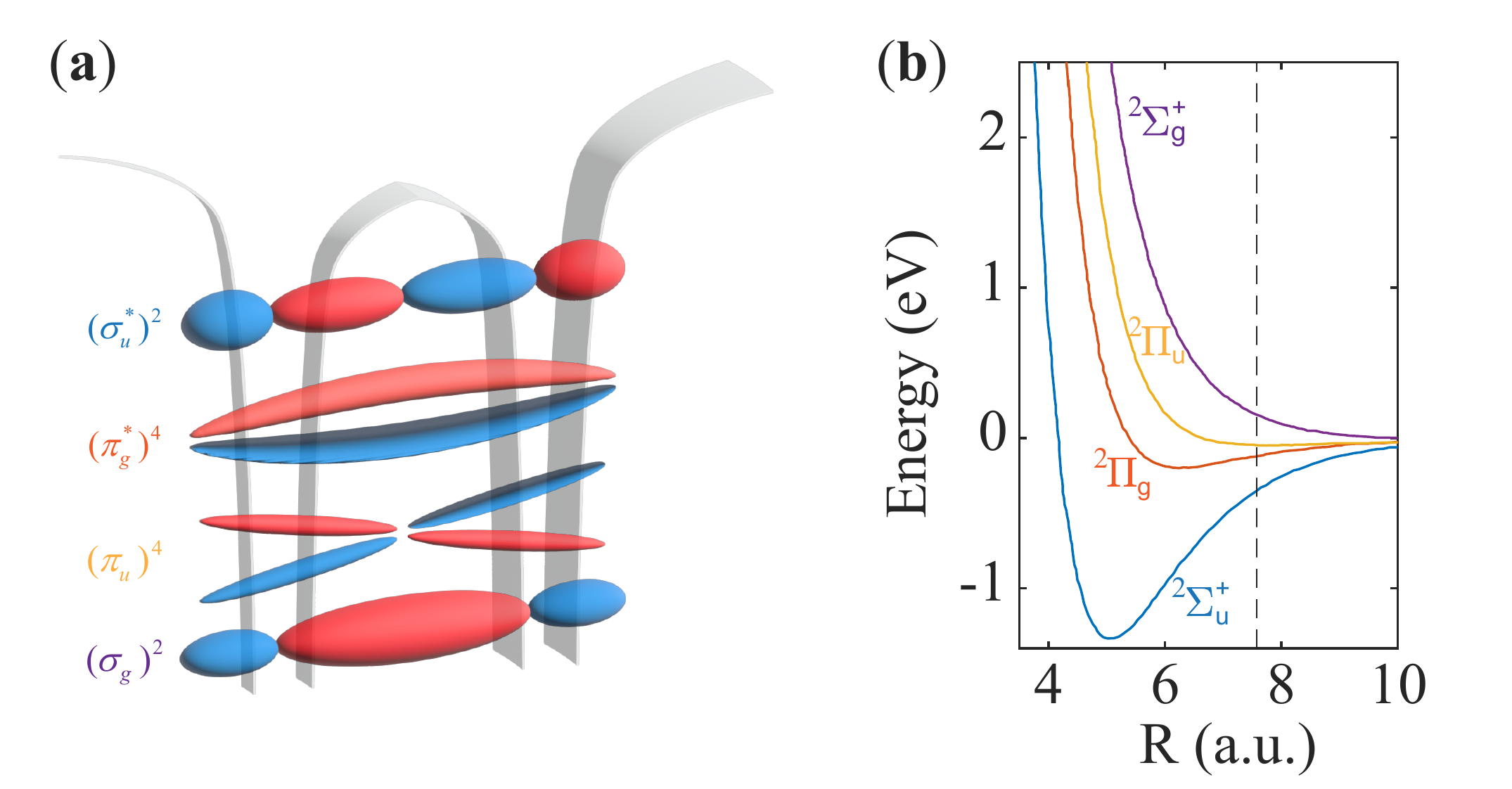}
\end{center}
{\textbf{Fig. 2.} Schematic illustration of the outermost valence molecular orbitals (\textbf{a}) for Kr$_2$ and the corresponding potential energy curves (\textbf{b}) for Kr$_2^+$, neglecting spin-orbit interaction for simplicity. In (b), the data is taken from Ref. \cite{kalus2003modelling} and the vertical dashed line indicates the equilibrium internuclear distance (7.578 a.u.) for the ground state of neutral Kr$_2$.}
\label{figure2}
\end{figure}

In Kr$_2$, the configuration of the outermost valence electrons is $(\sigma_g)^2(\pi_u)^4(\pi_g^*)^4(\sigma_u^*)^2$. The removal of one electron from one of these four molecular orbitals gives rise to the ionic states of $^2\Sigma_g^+$, $^2\Pi_u$, $^2\Pi_g$ and $^2\Sigma_u^+$, respectively, which are graphically illustrated in Fig. 2a. The different ionic states will have different nuclear dynamics after photoionization, resulting in different fragments. In Fig. 2b, we show the potential energy curves of the four ionic states as a function of the internuclear distance. $^2\Pi_g$ and $^2\Sigma_u^+$ states (corresponding to ionization of the anti-bonding orbitals) have a potential well, allowing the Kr$_2^+$ to remain bound. Thus, our experimental coincidence measurements, performed with the undissociated Kr$_2^+$, rule out the contributions from the other two states ($^2\Pi_u^+$ and $^2\Sigma_g$). In spite of this important simplification, one still needs to consider two ionic states of opposite parities. The parity of the molecular orbital controls the initial phase difference between the emitted electron wavepackets from the two centers. The gerade orbital launches wavepackets with a equal initial phase, whereas the wavepackets released from the two centers in the ungerade orbital have an initial $\pi$ phase shift. 

To demonstrate the parity effect in two-center interference on the photoionization delays, we first resort to a simple and intuitive model, i.e., we numerically solve the time-independent Schrödinger equation (TISE) for a 1D model potential with parameters chosen to closely resemble Kr$_2$. We used the double-well potential
\begin{equation}
 V(x)=V(x,-R)+V(x,R),
\end{equation}
where $R$=7.578 a.u. and the potential shape is given by
\begin{equation}
    V(x,x_0)=V_1 \frac{e^{-\frac{|x-x_0|}{\lambda}}}{\sqrt{(x-x_0)^2+s^2}} ,
\end{equation}
with $s$=1, $V_1$=-2.25 and $\lambda$=3. The ground state (gerade) and the first excited state (ungerade) of this double-well potential can be regarded as the molecular orbitals constructed by linear combination of two atomic $s$ orbitals, just like the case of H$_2$. The second (gerade) and third (ungerade) excited states are in analogy to the molecular orbitals constructed by two atomic $p$ orbitals, where the $\pi$ symmetry cannot be simulated by a 1D model. Therefore, in our calculations we use the second and third excited states as the initial state, and the choice of potential parameters ($s, V_1, \lambda$) gives the correct ionization potential ($I_{\rm p}$ = 14.0 eV) for the second excited state compared with the $I_{\rm p}$ of Kr. In Figure 3c, we illustrate the used double-well potential and the two normalized initial-state wavefunctions in coordinate space. Within time-independent perturbation theory, the electric transition matrix element is calculated from $\langle \psi_i | \hat d | \psi_f \rangle$, where $\psi_i$ is the initial state and $\psi_f$ is the final continuum state. The operator $\hat d = \hat r$ is the dipole operator in the length gauge. Here we only consider dipole transitions in which the initial and final states have opposite parities. The photoionization cross section is then given by
\begin{equation} \label{eq:CS}
    \sigma=\frac{4\pi^2\omega}{3c} |\langle \psi_i | \hat d | \psi_f \rangle|^2 ,
\end{equation} 
shown in Fig. 3a, where $\omega$ is the photon frequency and $c$ is the speed of light. The energy derivative of the phase shift (argument of transition matrix element),
\begin{equation} \label{eq:Wigner}
\tau_{\rm Wigner}=\frac{\partial {\rm Arg}( \langle \psi_i | \hat d | \psi_f \rangle)}{\partial E} ,
\end{equation}
gives the photoionization time delay shown in Figure 3b. In all panels, we use \textit{gerade} (in blue) and \textit{ungerade} (in red) labels according to the symmetry of the initial state.

For each initial state, we observe peaks in the time delay that correspond to pronounced minima in the cross section due to destructive interference. More importantly, the peaks of the gerade state corresponds to the minima of the ungerade state and vice versa, which demonstrates the initial $\pi$ phase shift between the two cases. We note that here the effect of two-center interference can be observed below 20~eV due to the very large internuclear distance of krypton dimer. In contrast, for the tightly bound molecules such as H$_2$ the energy range should cover up to several hundreds electron volts \cite{Ning2014a}, which is a challenge for experiments.

\begin{figure}
\begin{center}
\includegraphics[width=8.6cm]{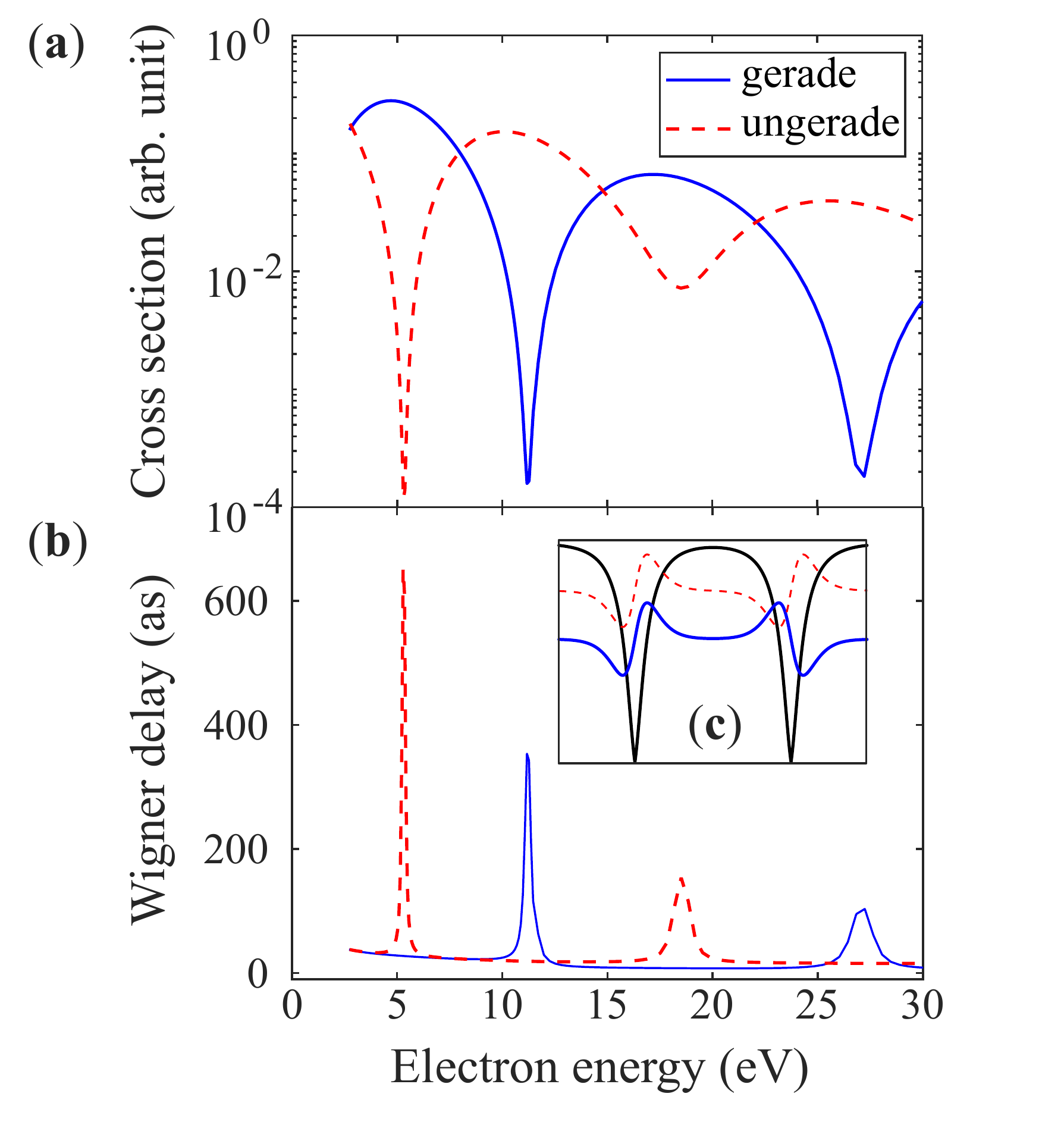}
\end{center}
{\textbf{Fig. 3.} (a) Photoionization cross section of the 1D system calculated with Eq. \eqref{eq:CS}. (b) Corresponding Wigner delay calculated with Eq. \eqref{eq:Wigner}. (c) Potential used in the calculations (black line) with the initial-state wavefunctions (blue and red lines). The ionization energies of the two states are 14.0 eV and 13.06 eV, respectively, and their vertical separation is added arbitrarily for clarity.}
\label{figure3}
\end{figure}

From the RABBIT spectrograms shown in Figure 1, we have extracted the phases of yield oscillations for the six sidebands using energy gating, followed by Fourier transformation. The uncertainty of the extracted sideband phase for each species (Kr and Kr$_2$) was determined by the A-over-B method \cite{ji2021quantitative} and that of their relative time delay was accordingly determined by the error propagation formula. In the Supplementary Material (SM), we illustrate the details of data analysis. Figure 4a displays the extracted relative ionization delays $\Delta \tau_{\rm Kr_2-Kr}$ between Kr$_2$ and Kr as a function of the sideband order or electron kinetic energy. The oscillation of the relative ionization delay around 0~as can clearly be seen. The amplitude of the oscillation is gradually damped with increasing electron kinetic energy. 

We further performed state-of-the-art photoionization calculations and extracted the photoionization delays for Kr$_2$ and Kr. The cross-section-weighted relative delay $\Delta \tau_{\rm Kr_2-Kr}$ is shown and compared with the experimental result in Fig. 4a, where good agreement between theory and experiment is achieved. These photoionization calculations were performed using the multichannel Schwinger configuration interaction (MCSCI) method \cite{Stratmann1995,Stratmann1996} to obtain the photoionization matrix elements and the formalism outlined in \cite{baykusheva17a} to obtain the photoionization delays. Here we used a single-center expansion with $l_{\rm{max}}$ = 200 to represent all bound and continuum functions. The initial bound state was the Hartree-Fock state computed using a correlation-consistent polarized valence triple zeta (cc-pVTZ) basis set \cite{Wilson1999} using MOLPRO \cite{Werner2012}. The ion states were then the frozen-core states created by removing one electron from the 4p orbitals. These four ionic states were all included in a close-coupling calculation and only the results of $^2\Sigma_{\rm{u}}^+$ and $^2\Pi_{\rm{g}}$ state were extracted and displayed. All calculations were computed with a fixed inter-nuclear distance of 7.578 a.u. (i.e. 4.01~${\rm{\AA}}$) \cite{Barker1974}.

The state-resolved relative ionization delays for $^2\Sigma_{\rm u}^+$ and $^2\Pi_{\rm g}$ states, that give rise to the undissociated dimer, are shown in Fig. 4b. Each of them oscillates with the electron energy and between them the anticyclic behavior can clearly be observed. Interestingly, the peak positions of the delays from gerade and ungerade wavefunctions roughly coincide with the peak positions resulting from the 1D TISE calculation, which is a further indication that the two-center interference is the cause for the oscillations observed in theory and experiment. Further we see that the averaged relative ionization delays in Fig. 4a are dominated by the delays of the $^2\Pi_{\rm g}$ state. The photoionization cross sections of $^2\Sigma_{\rm u}^+$ and $^2\Pi_{\rm g}$ are shown in SM. The cross section of $^2\Pi_{\rm g}$ is 2-4 times larger than that of $^2\Sigma_{\rm u}^+$. This is partially due to the fact that there are two degenerated $^2\Pi_{\rm g}$ states with $\Lambda$=+1 and -1 for the orbital-angular-momentum projection quantum number which need to be summed over for the cross section. As a result of the different cross sections, the interference from gerade and ungerade wavefunctions does not cancel out completely, which results in an observable oscillation in the relative ionization delay. We note that the photoionization cross sections also display anti-cyclic oscillations in their amplitude (see SM), which match the oscillations in the state-resolved delays displayed in Fig. 4b. Comparing these accurate calculations to the solution of the time-independent Schrödinger equation of our 1D model potential, we conclude that the oscillations observed in the experiment and predicted by the MCSCI calculations are being caused by the interference of electron wavepackets departing or scattering from the two atoms in Kr$_2$. The large positive and negative delays in the energy range below 5 eV, in experiment and theory in Fig. 4, may suggest that two-center interference between the photoelectron wave departing from one site and the diffracted wave from the other site contributes significantly to the enhancement of the time delays. To evaluate the contribution from this diffraction effect, additional MCSCI calculations have been carried out with initial states localized to a single Kr atom, in which case the photoelectron is emitted from one Kr atom only and is being diffracted from the other Kr atom: the calculated time delays (not shown) are almost identical to the results in Fig. 4(b), confirming that the observed oscillatory structure in the time delay mainly comes from the two center interference between the photoelectron wave emitted from one Kr atom and the diffracted wave from the other Kr atom.

\begin{figure}
\begin{center}
\includegraphics[width=8.6cm]{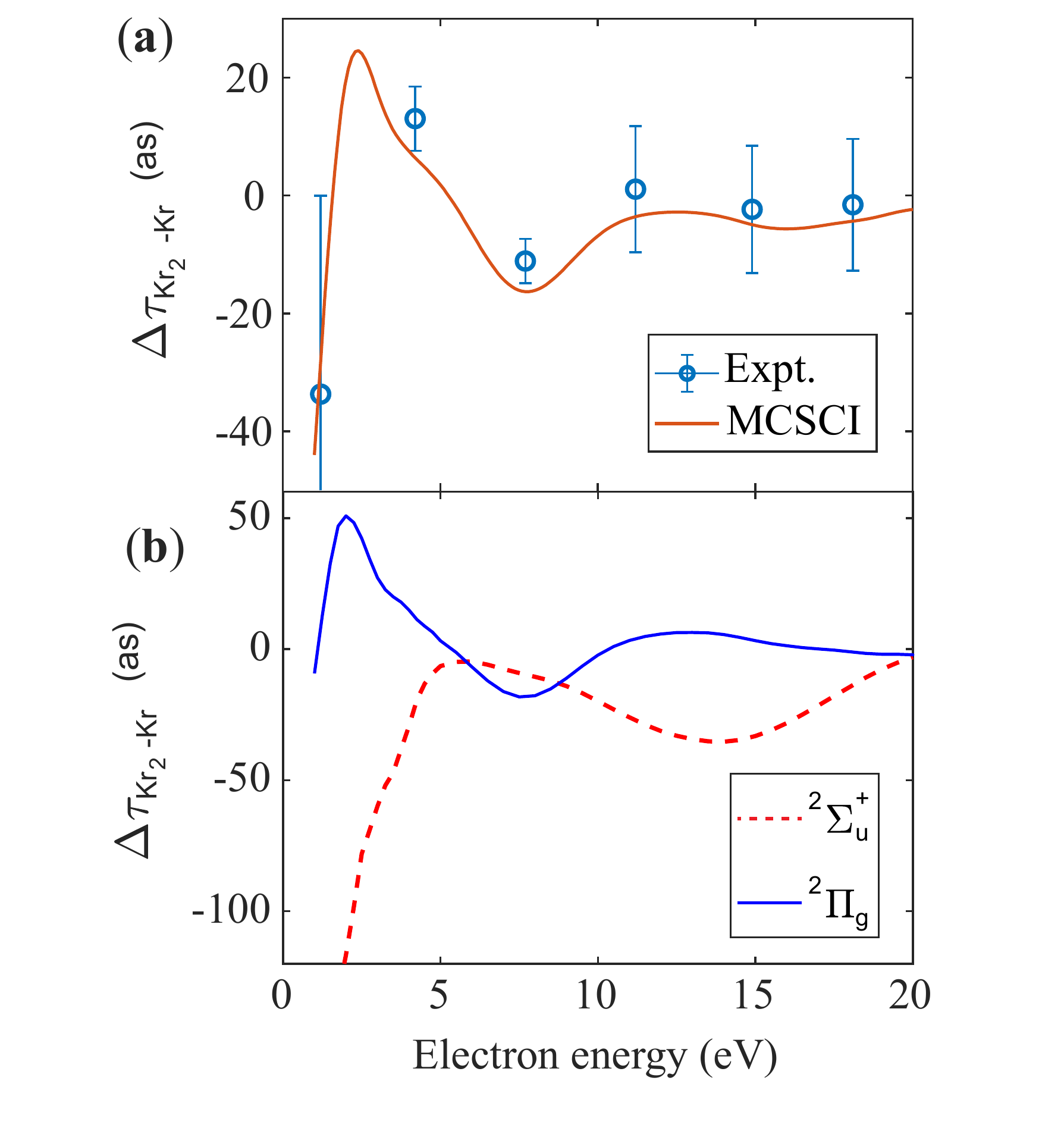}
\end{center}
{\textbf{Fig. 4.} (a)  ionization delays $\Delta\tau_{\rm Kr_2-Kr}$ from experimental data (open circles), where values show the phase differences extracted from Figure 1(b-c) and the error bars represent the uncertainty of the phase extraction. The relative ionization delays from the MCSCI calculations are shown as a solid line. Theory for Kr$_2$ includes the cross section weighted delays for the two bound states of Kr$_2^+$, i.e., $^2\Sigma_{\rm u}^+$ and $^2\Pi_{\rm g}$. (b) State-resolved relative delays from the MCSCI calculations for the $^2\Sigma_{\rm u}^+$ and $\Pi_{\rm g}$ states. Note that the solid line in (a) is the cross section weighted average of the state resolved delays in (b). For both experiment and theory the electron emission angles range from $\Theta_{\rm Lab}=0-25^\circ$ with respect to the XUV polarization, and the average of the molecular axis in the lab frame is included in calculations.}
\label{fiure4}
\end{figure}

In conclusion we have demonstrated the manifestations of two-center quantum interference in the ionization time delays of a diatomic homonuclear molecule. We have done so in a fundamental theoretical manner by solving the time-independent Schrödinger equation of a 1D double-well potential and experimentally by measuring the relative ionization delays between Kr and Kr$_2$, where the measurement result is quantitatively supported by state-of-the-art quantum scattering calculations. These results show that two-center-interference effects can be observed in attosecond photoionization delays. Such effects can be expected to be observed in many other systems as well, provided that the internuclear separation is sufficiently large or the electron-kinetic energy if sufficiently high, to fulfill the interference condition. Our results also show that the opposite modulations of initial states of opposite parity tend to cancel the signatures of two-center interference, which explains why the observed effects are relatively small. In cases where the energy intervals corresponding to ionization from initial states of different parity are resolvable, correspondingly larger effects can be expected. This is the case, in particular, in lighter diatomic molecules with shorter internuclear separations. However, two-center-interference effects can also be expected in larger systems consisting of two identical subunits, such as biphenyl and its derivatives or two-center metal complexes, be it in the gas or liquid phase \cite{jordan20a}. Our results therefore pave the way to the investigation of a broad variety of quantum-interference effects in attosecond chronoscopy.

We thank A. Schneider and M. Seiler for their technical support. We gratefully acknowledge funding from the Swiss National Science Foundation (SNSF) through projects 206021\_170775 and 200021\_172946, as well as ERC Project No. 772797. Work by MH and DJ was supported by European Union’s Horizon 2020 program under MCSA Grant No 801459, FP-RESOMUS. Work by RRL was supported by the U.S. Department of Energy (DOE), Office of Science, Basic Energy Sciences (BES) under Contract DE-AC02-05CH1123.

\bibliography{attobib,MyThesisBib}

\end{document}